\documentclass[a4paper,twocolumn,superscriptaddress,accepted=2024-10-28]{quantumarticle}
\pdfoutput=1

\usepackage[utf8]{inputenc}
\usepackage[american]{babel}
\usepackage[T1]{fontenc}
\usepackage[pdftex]{graphicx}  
\usepackage{xcolor}
\usepackage{dcolumn}
\usepackage{bm}
\usepackage{dsfont}
\usepackage{amsmath,amsthm,amssymb}
\usepackage{braket}
\usepackage{wrapfig}
\usepackage{mathtools}
\usepackage{float}
\usepackage{physics}
\usepackage[normalem]{ulem}

\usepackage{hyperref}
\hypersetup{colorlinks,bookmarksopen,bookmarksnumbered,citecolor=red,linkcolor=red,pdfstartview=false,urlcolor=red}
\usepackage[numbers,sort&compress]{natbib}
\newcommand{\daga}{^\dag}

\begin{document}
\title{Enhanced Entanglement in the Measurement-Altered Quantum Ising Chain}

\author{Alessio Paviglianiti}
\affiliation{International School for Advanced Studies (SISSA), via Bonomea 265, 34136 Trieste, Italy}
\author{Xhek Turkeshi}
\affiliation{Institut für Theoretische Physik, Universität zu Köln, Zülpicher Strasse 77, 50937 Köln, Germany}
\affiliation{JEIP, UAR 3573 CNRS, Coll\`{e}ge de France, PSL Research University, 11 Place Marcelin Berthelot, 75321 Paris Cedex 05, France}
\author{Marco Schir\`o}
\affiliation{JEIP, UAR 3573 CNRS, Coll\`{e}ge de France, PSL Research University, 11 Place Marcelin Berthelot, 75321 Paris Cedex 05, France}
\author{Alessandro Silva}
\affiliation{International School for Advanced Studies (SISSA), via Bonomea 265, 34136 Trieste, Italy}

\begin{abstract}
Understanding the influence of measurements on the properties of many-body systems is a fundamental problem in quantum mechanics and for quantum technologies.
This paper explores how a finite density of stochastic local measurement modifies a given state's entanglement structure. 
Considering various measurement protocols, we explore the typical quantum correlations of their associated projected ensembles arising from the ground state of the quantum Ising model. 
Using large-scale numerical simulations, we demonstrate substantial differences among inequivalent measurement protocols. 
Surprisingly, we observe that forced on-site measurements can enhance both bipartite and multipartite entanglement. 
We present a phenomenological toy model and perturbative calculations to analytically support these results. 
Furthermore, we extend these considerations to the non-Hermitian Ising model, naturally arising in optically monitored systems, and we show that its qualitative entanglement features are not altered by a finite density of projective measurements.
Overall, these results reveal a complex phenomenology where local quantum measurements do not simply disentangle degrees of freedom, but may actually strengthen the entanglement in the system.
\end{abstract}
\maketitle

\section{Introduction} 
Measurements lie at the core of quantum mechanics. Observing the properties of a system inevitably perturbs it~\cite{Shankar:102017,gisin1992thequantumstate}. 
Investigating the role and consequences of measurements has a long tradition in multiple branches of quantum physics, including quantum information theory~\cite{breuer2002thetheoryof,jacobs2014quantummeasurementtheory}, condensed matter~\cite{Daley2014}, and quantum optics~\cite{carmichael1999statisticalmethodsin,gardiner2004quantumnoise,wiseman2009quantummeasurementand}.
In a wide range of problems, the only role measurements play is extracting information from a system, whereas the post-observation state is often disregarded. Recently, this paradigm has been turned upside down in the study of monitored many-body systems, where the system's unitary dynamics is interspersed with controlled local probing of a measurement apparatus. 
In this context, measurements constitute an active ingredient of the dynamics, as they alter the state and lead to fascinating non-unitary phenomena, hallmarked by measurement-induced phase transitions~\cite{fisher2023randomquantumcircuits,cao2019entanglementina,li2018quantumzenoeffect,li2019measurementdrivenentanglement,skinner2019measurementinducedphase,zabalo2020criticalpropertiesof,vasseur2019entanglementtransitionsfrom,barratt2022fieldtheoryof,choi2020quantumerrorcorrection,bao2020theoryofthe,gullans2020dynamicalpurificationphase,gullans2020scalableprobesof,agrawal2022entanglmentandchargesharpening,block2022measurementinducedtransition,sharma2022measurementinducedcriticality,muller2022measurementinduceddark,minato2022fateofmeasurementinduced,sierant2022dissipativefloquetdynamics,passarelli2023postselectionfree,sierant2022measurementinducedphase,klocke2023majorana,lunt2021measurementinducedcriticality,turkeshi2020measurementinducedcriticality,sierant2022universalbehaviorbeyond,nahum2021measurementandentanglement,szyniszewski2020universalityofentanglement,potter2022quantumsciencesandtechnology,lunt2022quantumsciencesandtechnology,russomanno2023}.
The interest in monitored models is on the rise, especially in view of the advent of noisy intermediate-scale quantum devices~\cite{preskill2018quantumcomputingin,Fraxanet2023,ferris2022quantumsimulationon}.

More recently, the study of measurement-altered phases of matter has also been pursued in a static scenario, addressing the question of how a finite density of measurements affects the properties of a given quantum state~\cite{garratt2022measurements,weinstein2023nonlocality,lee2023quantum,zhu2022nishimoris,yang2023entanglement,sun2023new,su2023higherform,murciano2023measurementaltered}. 
In detail, Refs.~\cite{garratt2022measurements,weinstein2023nonlocality,lee2023quantum,yang2023entanglement,sun2023new,murciano2023measurementaltered} investigate measurement-altered \emph{critical} ground states, focusing on correlation functions and on the entanglement properties of Luttinger liquids and free fermionic states. Compared to the dynamical setup of monitored systems~\cite{alberton2021entanglementtransitionin,buchhold2021effectivetheoryfor,ladewig2022monitoredopenfermion,turkeshi2023density,turkeshi2022enhancedentanglementnegativity,turkeshi2021measurementinducedentanglement,turkeshi2023entanglementandcorrelation,turkeshi2022entanglementtransitionsfrom,paviglianiti2023multipartiteentanglementin,zerba2023measurement,piccitto2022entanglementtransitionsin,tirrito2023fullcountingstatistics,coppola2022growthofentanglement,fava2023nonlinearsigmamodels,jian2023measurementinducedentanglement,jian2021measurementinducedphase,poboiko2023theoryoffree,loio2023purificationtimescalesin}, this approach reveals equilibrium aspects of the model, and constitutes a natural framework to explore the effect of measurements in many-body systems.

This work investigates the robustness of quantum correlations and of the entanglement properties of critical and \emph{non-critical} states to a finite density $p\in[0,1]$ of measurements. 
We consider the ground states of the quantum Ising chain for multiple parameter choices, allowing us to probe different phases. 
In our analysis, we focus on three entanglement witnesses of bipartite and multipartite quantum correlations: the entanglement entropy (EE), the quantum Fisher information (QFI), and the two-body fermionic negativity (FN). 
We explore how different realizations of the measurement outcomes induce inequivalent phenomenology and distinct entanglement behavior. Concretely, we consider Born rule and forced measurements, the former manifesting the intrinsic stochasticity of quantum mechanics, the latter forcing projections along a pre-determined direction.

Our main result is that, contrary to common belief, local measurements do not necessarily reduce entanglement, but can even enhance it in certain situations. 
This undermines the heuristic interpretation of measurement-induced phase transitions as a competition between entanglement-increasing unitary dynamics and correlation-destroying monitoring. 
We develop a theoretical argument to understand the effect, and build a toy model that captures the physics of this mechanism, which is deeply tied to the monogamy of entanglement.
Finally, we investigate a measurement-altered non-Hermitian Ising chain. This approach aims at establishing to what extent non-Hermitian models can describe accurately the physics of measurement-induced phase transitions~\cite{biella2021manybodyquantumzeno,paviglianiti2023multipartiteentanglementin,zerba2023measurement}, and we reveal that the scaling properties of entanglement witnesses are unchanged by our protocol. Our findings indicate that any discrepancy between the non-Hermitian description and the actual physics is dynamical, i.e., it is a consequence of the combined effect of dynamics and measurements.

The rest of this manuscript is organized as follows. 
First, we define the framework of our study in Sec.~\ref{sec:maic}, as well as the observables we consider. 
Afterward, we showcase numerical simulations in Sec.~\ref{sec:numerics}, and explain them using perturbation theory. Motivated by these results, we develop theoretical approaches to explain the phenomenology in Sec.~\ref{sec:network}, where we introduce a toy model of an entanglement network that describes the enhancement of correlations. In Sec.~\ref{sec:mipt}, we extend our investigation to the measurement-altered non-Hermitian Ising chain, and we characterize the impact of forced projections on the non-Hermitian Hamiltonian description of certain models of measurement-induced phase transitions. A conclusive discussion of the results obtained in given in Sec.~\ref{sec:conclusion}. 

\section{Model and Entanglement Witnesses}
\label{sec:maic}
We are interested in studying the typical properties of a given state $|\Psi_0\rangle$ of $L$ spins subject to a finite density of local measurements. The initial state we consider is the ground state of the quantum Ising chain
\begin{equation}\label{ising}
    \hat{H} = -\sum_j\hat{\sigma}_j^x\hat{\sigma}_{j+1}^x -h\sum_j \hat{\sigma}_j^z,
\end{equation}
with periodic boundary conditions. The perturbation of this state with measurements has already been considered in Refs.~\cite{murciano2023measurementaltered,weinstein2023nonlocality} in the critical case, here we extend the study to generic choices of $h$, including non-Hermitian setups [cf. Sec.~\ref{sec:mipt}].

In the following, we assume that each lattice site has a finite probability $p$ of being addressed by measurement. To this purpose, for each site $j$ we introduce a discrete random variable $m_j$ such that $m_j=0$ with probability $1-p$ and $m_j\neq 0$ with probability $p$, corresponding to a spin being measured or not, respectively. On average, $pL$ sites are measured. Whenever $m_j\neq 0$, it can assume the values $m_j = \pm 1$, which indicate the possible measurement outcomes. Specifically, we assume that the state is projected locally onto either $\ket{1_j}$ or $\ket{-1_j}$, which we assume to be the eigenstates of $\hat{\sigma}^z_j$~\footnote{Throughout this paper $\hat{\sigma}^\alpha$ with $\alpha=x,y,z$ are the Pauli matrices.}, defined as 
$\hat{\sigma}^z_j\ket{\pm 1} = \pm \ket{\pm 1}$. Any post-observation state can be characterized by assigning the string $m=(m_1,m_2,\dots, m_L)$, and it is given by
\begin{equation}
    |\Psi_m\rangle = \frac{\hat{\Pi}^{(m)} |\Psi_0\rangle}{||\hat{\Pi}^{(m)} |\Psi_0\rangle||},\quad \hat{\Pi}^{(m)} = \prod_{j=1}^L\hat{\pi}_j^{(m_j)},
    \label{eq:proj}
\end{equation}
where $\hat{\pi}_j^{(0)} = \mathds{1}_j$, and $\hat{\pi}_j^{(\pm 1)} = \ket{\pm 1_j}\bra{\pm 1_j}$.
When $m_j\neq 0$, the choice of the local post-measurement state $\ket{m_j}$ depends on the specific measurement protocol we adopt, which is one of the following:
\begin{enumerate}
    \item[(i)] forced up measurements ($\mathcal{M}_\mathrm{up}$), where the post-measurement state is $\ket{1_j};$
    \item[(ii)] forced down measurements ($\mathcal{M}_\mathrm{down}$), where the post-measurement state is $\ket{-1_j};$
    \item[(iii)] Born rule measurements ($\mathcal{M}_\mathrm{Born}$), where the measurement outcomes are extracted randomly according to the Born rule, i.e., $\ket{m_j}$ is picked with probability $|\langle m_j|\Psi\rangle|^2$.
\end{enumerate}
We point out that (i) and (ii) do not constitute proper quantum measurements, but rather they are exponentially unlikely post-selected outcomes of (iii). Still, certain quantum-jump unravelings of Lindbladian dynamics can actually give rise to projections in a fixed direction, together with an effective non-Hermitian Hamiltonian evolution~\cite{paviglianiti2023multipartiteentanglementin,zerba2023measurement,piccitto2022entanglementtransitionsin,maki2023montecarlo}. In this context, the investigation of forced jumps can be used to characterize the stability against sudden projections of the non-Hermitian physics produced by the Hamiltonian alone, which has been investigated in models of measurement-induced phase transitions [see Sec.~\ref{sec:mipt}]. 
Moreover, as we demonstrate below, forced projections can affect entanglement in unexpected ways, elucidating therefore how quantum jumps perturb a state in general. Their investigation thus provides information on the behavior of atypical random realizations.

We frame our discussion in a unified perspective using the concept of \emph{projective ensembles}~\cite{ho2022exactemergentquantum,Claeys2022emergentquantum,Ippoliti2022solvablemodelofdeep,ippoliti2022dynamicalpurificationand,lucas2022generalizeddeepthermalization,lydzba2022,Choi1,choi2} by introducing the outcome probability distributions
\begin{subequations}\label{prob_densities}
    \begin{equation}\label{prob_density_up}
        \mathcal{P}_{\mathcal{M}_\mathrm{up}}(m) = p^{\sum_j |m_j|} (1-p)^{\sum_j(1-|m_j|)} \prod_j (1-\delta_{m_j,-1}),\\
    \end{equation}
    \begin{equation}
        \mathcal{P}_{\mathcal{M}_\mathrm{down}}(m) = p^{\sum_j |m_j|} (1-p)^{\sum_j(1-|m_j|)} \prod_j (1-\delta_{m_j,1}),\\
    \end{equation}
    \begin{equation}
        \mathcal{P}_{\mathcal{M}_\mathrm{Born}}(m) = p^{\sum_j |m_j|} (1-p)^{\sum_j(1-|m_j|)} \bra{\Psi_0}\hat{\Pi}_{m} \ket{\Psi_0}
    \end{equation}
\end{subequations}
for the different protocols.
A measurement-altered system is described by the ensemble $\mathcal{E}_{\mathcal{M}}(|\Psi_0\rangle) =\left\lbrace \left(\mathcal{P}_{\mathcal{M}}(m),|\Psi_m\rangle\right)\right\rbrace$. The statistical properties of $\mathcal{E}_{\mathcal{M}}(|\Psi_0\rangle)$ contain more information than the average state $\hat{\rho}_1 = \sum_m \mathcal{P}_\mathcal{M}(m) |\Psi_m\rangle\langle \Psi_m|$. Indeed, while $\hat{\rho}_1$ perfectly describes all linear functions, averaged over the distribution $\mathcal{P}_{\mathcal{M}}(m)$, of $|\Psi_m\rangle \langle\Psi_m|$, it does not capture the typical properties of non-linear functionals of the density matrix. In fact, such functionals include effects of the $k$-replicated desnity matrix
\begin{equation}
    \hat{\rho}_k = \sum_m\mathcal{P}_\mathcal{M}(m) (|\Psi_m\rangle\langle \Psi_m|)^{\otimes k},
\end{equation}
and reveal non-trivial beyond-average quantum correlations.

Entanglement measures and witnesses are examples of non-linear quantities, and they are the main focus of the present manuscript. Specifically, we consider the entanglement entropy and the quantum Fisher information as functions of the measurement density $p$. Furthermore, we also investigate the fermionic negativity  of spin pairs, from which we can extract the typical length scale of quantum correlations. Given our choice of the initial state and of the measurement operators, the problem can be studied efficiently using Gaussian states, reviewed for instance by Ref.~\cite{tagliacozzo}. This not only allows us to explore large system sizes, but makes the calculation of the entanglement witnesses efficient. A detailed description of our Gaussian state implementation is presented in App.~\ref{a:implementation}.

The EE between a compact interval $A$ of $\ell$ spins and the complementary subsystem $\overline{A}$ is defined as~\cite{Laflorencie_2016,amico2008entanglementinmanybody,calabrese2004entanglemententropyand,calabrese2009entanglemententropyand}
\begin{equation}
    S_\ell = -\Tr\left(\hat{\rho}_A\ln\hat{\rho}_A\right)
\end{equation}
where $\hat{\rho}_A = \Tr_{\;\overline{A}}\ket{\Psi}\bra{\Psi}$ is the reduced density matrix. The entanglement entropy of a Gaussian state can be computed efficiently by diagonalizing the covariance matrix of subsystem $A$, as described for instance in Refs.~\cite{tagliacozzo,mbeng2024}.

The QFI of a pure state $|\Psi\rangle$ with respect to a given observable $\hat{O}$ is given by~\cite{pezze2018quantummetrologywith,pappalardi2017multipartiteentanglementafter,pappalardi2018scramblingandentanglement,pappalardi2022extensivemultipartiteentanglement,brenes2020multipartitenetanglementstructure,paviglianiti2023multipartiteentanglementin}
\begin{equation}
    F_Q[\hat{O}] = 4\left(\langle\Psi|\hat{O}^2|\Psi\rangle-\langle\Psi|\hat{O}|\Psi\rangle^2\right).
\end{equation}
The QFI is related to multipartite entanglement when one studies its behavior over the set of operators $\hat{O}[\{\mathbf{n}_j\}] = \sum_j\mathbf{n}_j\cdot\hat{\boldsymbol{\sigma}}_j/2$, with $\norm{\mathbf{n}_j}=1$.
Using this expression for $\hat{O}$, the QFI is rewritten as
\begin{equation}
    F_Q[\hat{O}[\{\mathbf{n}_j\}]]=\sum_{\alpha,\beta} \sum_{i,j}n_i^\alpha n_j^\beta C^{\alpha,\beta}_{i,j}
\end{equation}
in terms of the connected spin-spin correlators $C^{\alpha,\beta}_{i,j} = \langle\hat{\sigma}^\alpha_i\hat{\sigma}^\beta_j\rangle-\langle\hat{\sigma}^\alpha_i\rangle\langle\hat{\sigma}^\beta_j\rangle$~\cite{caianiello1952,barouch1971}. As proved in Ref.~\cite{pezze2018quantummetrologywith}, a QFI density $f_Q=F_Q/L$ strictly greater than some divider $k$ of $L$ indicates the presence of $(k+1)$-partite entanglement in the quantum state. A lower bound on multipartite entanglement is then obtained as the maximal QFI density $f_Q^\mathrm{max}=\max_{\{\mathbf{n}_j\}}f_Q[\hat{O}[\{\mathbf{n}_j\}]]$. In this work, the maximal QFI is evaluated as discussed in Ref.~\cite{paviglianiti2023multipartiteentanglementin} by computing the spin-spin correlators and performing the maximization with an annealing algorithm.

Finally, the FN is a measure of bipartite entanglement similar to the entropy, but with the advantage of applying also to mixed states~\cite{shapourian2017partial,eisert2018entanglement,neven2021symmetry,vitale2022symmetryresolved,shapourian2019twisted,turkeshi2020negativity,ruggiero2022quantuminformationspreading,sara1,sara2,alba1,alba2,rottoli2023,rottoli2023bis}, differently for instance from the quantum mutual information~\cite{groisman2005,amico2008entanglementinmanybody}. In particular, given a tripartition of the system in distinct subsets $A_1$, $A_2$, and $\overline{A_1 \cup A_2}$, the negativity allows us to investigate the entanglement between $A_1$ and $A_2$ alone. In the following, we assume that $A_1$ and $A_2$ are single spins at positions $i$ and $j$, and we are interested in how their negativity decays with the distance $|j-i|$. First, we define $A = A_1 \cup A_2$, and we consider the reduced density matrix $\hat{\rho}_A = \Tr_{\;\overline{A}}\ket{\Psi}\bra{\Psi}$. The FN is then given by
\begin{equation}
    E^f_{i,j} = \ln \Tr\left|\hat{\rho}_A^{\Tilde{T}_1}\right|,
\end{equation}
where $\hat{\rho}_A^{\Tilde{T}_1}$ is the twisted partial transpose of $\hat{\rho}_A$ with respect to subsystem $A_1$~\footnote{We highlight that the FN does not coincide with the regular negativity of bosonic systems. A detailed discussion on why the FN was introduced is presented in Refs.~\cite{shapourian2017partial,sara1}}. The numerical calculation of the negativity is very similar to the entropy, and is described in detail in Refs.~\cite{shapourian2017partial,alba1,ruggiero2022quantuminformationspreading}.

\section{Numerical phenomenology and perturbative explanation}
\label{sec:numerics}

One of the main results of the present work is the demonstration that measurements do not only degrade entanglement, but might actually strengthen it. In this section, we show this effect using numerical simulations, and develop a qualitative theoretical explanation based on perturbation theory. First, let us illustrate the basic concept with a simple example. Consider the state 
\begin{equation}
\begin{split}
    |\phi\rangle &= \frac{1}{\sqrt{6}} (2|-1_A,1_B,1_C\rangle + \\ &\qquad + |1_A,-1_B,1_C\rangle + |1_A,1_B,-1_C\rangle) .
\end{split}
\end{equation}
The entanglement entropy between $C$ and the remainder is $S_C\approx 0.62$. Now, we perform a projective measurement on $A$. If the outcome is $\ket{-1_A}$, the post-measurement state is a product state $|\phi'_{-1}\rangle = |1_B 1_C\rangle$ with no entanglement, i.e., $\left.S'_C\right|_{-1} = 0$. 
In contrast, if the outcome is $\ket{1_A}$, the post-measurement state is the Bell pair $|\phi_1'\rangle = (|1_B -1_C\rangle + |-1_B 1_C\rangle)/\sqrt{2}$ and the post-observation entropy is $\left.S'_C\right|_{+1} = \ln 2 \approx 0.69$. Therefore, depending on the outcome, the measurement of $A$ can either destroy or enhance entanglement between $B$ and $C$. This phenomenon is referred to as measurement-induced entanglement~\cite{lin2023,cheng2024,zhang2024}. In the following, we investigate in which cases this effect appears in a many-body context.

We report numerical results for the quantities introduced in Sec.~\ref{sec:maic} using different measurement densities $p$ and various protocols. In particular, for ${\mathcal{Q}=S_\ell,\ F_Q,\ E^{f}_{i,j}}$ we study the ensemble averages $\overline{Q}$. The simulations are performed using Gaussian state techniques, which allow us to explore large system sizes up to $L=1024$ lattice sites. The numerical results are obtained by performing Monte Carlo sampling of the ensembles $\mathcal{E}_\mathcal{M}(|\Psi_0\rangle)$, using the probabilities discussed in Sec.~\ref{sec:maic}. We employ different choices of transverse field $h$ to probe the two phases of the Ising chain, as well as the critical point $h=1$.

As anticipated, at the critical point $h=1$, the model has been previously investigated in Ref.~\cite{weinstein2023nonlocality,lee2023quantum,sara1,yang2023entanglement}. These works demonstrate that the critical properties present crossover effects, with an effective central charge renormalized for forced up measurements and unaltered in the Born rule case.
In this manuscript, we extend these considerations also to the off-critical phase, and consider forced down measurements as well.
Our main result is that the paramagnetic phase reveals an entanglement enhancement for the forced measurements $\mathcal{M}_\mathrm{down}$, while a trivial crossover effect is observed for $\mathcal{M}_\mathrm{up/Born}$. 
Instead, the ferromagnetic phase shows robustness to measurements, irrespective of the measurement-altering protocol. 

\begin{figure*}[t!]
    \centering
    \includegraphics[width=\textwidth]{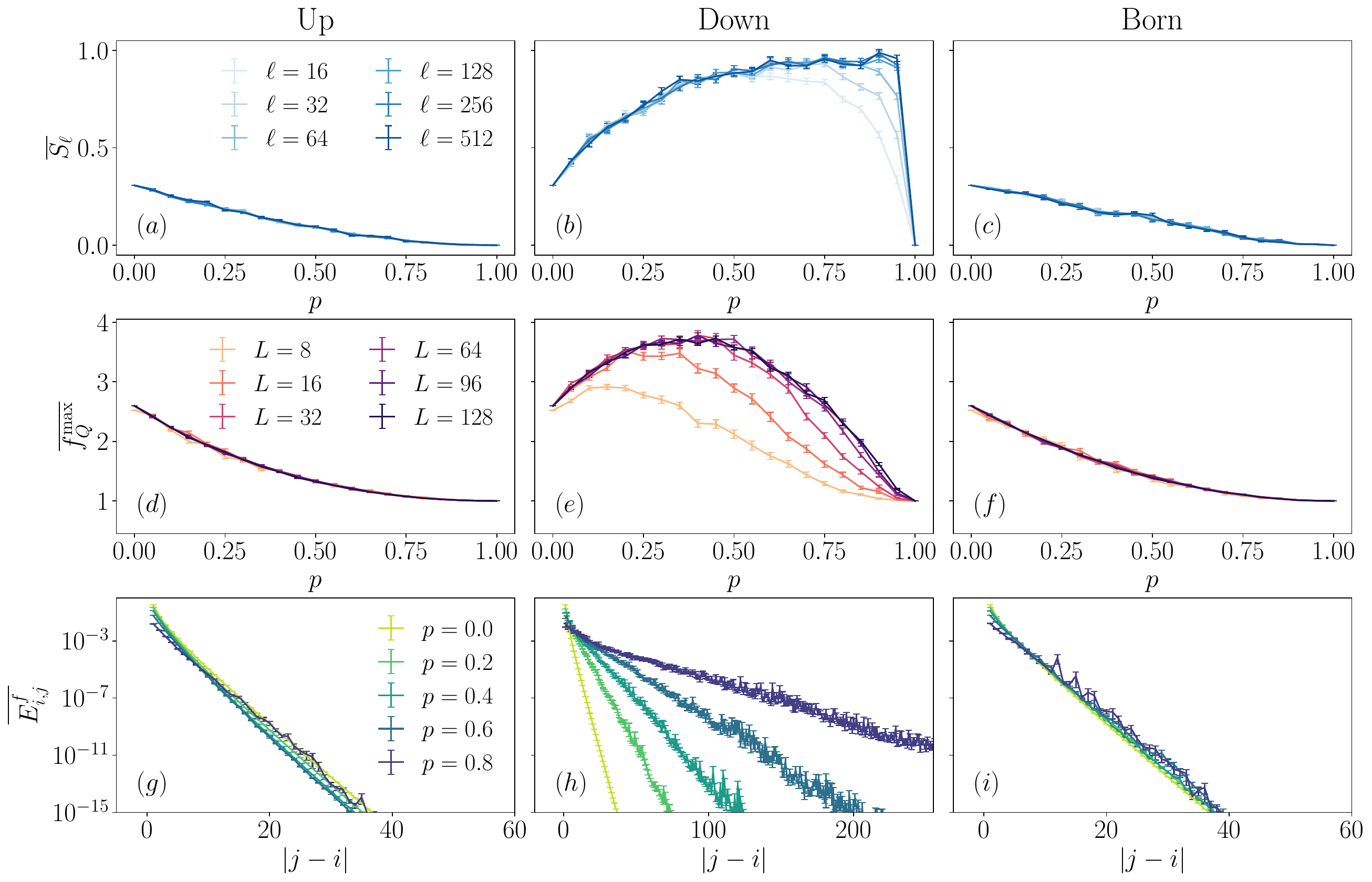}
    \caption{(a-c) EE, (d-f) maximal QFI density, and (g-i) pairwise FN of the perturbed GS of the quantum Ising chain with $h=1.5$. Left column: protocol $\mathcal{M}_\mathrm{up}$. Center column: protocol $\mathcal{M}_\mathrm{down}$. Right column: protocol $\mathcal{M}_\mathrm{Born}$. We adopt $L=1024$ for the EE and $L=512$ for the FN.}
    \label{fig:disordered}
\end{figure*}

\subsection{Disordered phase $h>1$}
We begin our analysis by considering the paramagnetic disordered phase at ${h>1}$. Fig.~\ref{fig:disordered} summarizes our results for the representative choice of ${h=1.5}$. 
In Fig.~\ref{fig:disordered} (a-c), we report the average EE for varying $p$ at different subsystem sizes. The protocols $\mathcal{M}_\mathrm{Born}$ and $\mathcal{M}_\mathrm{up}$ yield a similar size-independent behavior, monotonously decreasing with $p$. For Born measurements this is expected, as any entanglement monotone cannot increase on average under local operations~\cite{vidal2002,plenio2006}. In contrast, we find that forced down projections result in larger entanglement as compared to the unperturbed state. 
In particular, $S_\ell$ develops a peak as a function of $p$, migrating toward $p=1$ as $\ell$ is increased. This suggests that in the limit of $\ell\to\infty$ the entanglement entropy as a function of $p$ tends to a limiting curve that saturates to a constant for $p\to 1$. Our numerics suggests that $\lim_{p\to 1}\lim_{\ell\to\infty}S_\ell$ is independent of $\ell$, which implies that the enhanced entangled state still follows an area law.

Next, we analyze multipartite entanglement in Fig.~\ref{fig:disordered} (d-f). 
The average QFI density reveals a behavior similar to that of the EE: for the protocols $\mathcal{M}_\mathrm{up/Born}$, the QFI decreases when varying $p$, while forced down measurements yield enhanced QFI, with a peak value, saturated at large $L$, of $\overline{f^\mathrm{max}_Q}\simeq 4$. 
The maximal QFI is attained at around $p\simeq 0.45$, demonstrating that the measurements affect inequivalently multipartite and bipartite quantum correlations. The saturating behavior of $f_Q$ reveals bounded multipartiteness of entanglement.

Lastly, in Fig.~\ref{fig:disordered} (g-i) we report the average pairwise FN $\overline{E_{i,j}^f}$ as a function of the distance $|j-i|$ for different measurement densities.
The overall envelope of the decay is exponential, which is consistent with the presence of a finite length scale in the system, as demonstrated by the area law entanglement previously discussed. Importantly, for $\mathcal{M}_\mathrm{up/Born}$ the decay rate is independent of the measurement rate. Instead, for $\mathcal{M}_\mathrm{down}$, the negativity features a renormalized length scale, which grows as the measurement density $p$ increases. 
This reveals that the range of quantum correlations is effectively extended by the local measurements for $\mathcal{M}_\mathrm{down}$, whereas for $\mathcal{M}_\mathrm{up/Born}$ it remains unaffected. In Sec.~\ref{sec:network}, we develop a toy model to explain this effect qualitatively.

We stress that entanglement enhancement is a consequence of the monitoring alone, and no form of feedback is required. It is well known that many state-preparation protocols use measurements to realize long-range entangled states, but these techniques also rely on entangling unitary gates, often conditioned to the measurement outcomes. Moreover, these paradigms usually operate on special simple states, whereas our observation of enhancement applies to complex many-body states.

The observed phenomenology can be understood by considering the structure of the initial state $|\Psi_0\rangle$ within perturbation theory. For $h\gg 1$, we expand the ground state of the quantum Ising chain of Eq.~\eqref{ising} in powers of $h^{-1}$ starting from the ground state for $h\to\infty$, which is simply the product state $\ket{1_1\dots1_L}$. Throughout the following calculations, the normalization of the state is disregarded, as it is irrelevant for the purposes of our discussion. For later convenience, let us label this state as $\ket{vac}$. Perturbative corrections consist of states with local defects, i.e., spins opposite orientations with respect to the rest of the chain. We denote a state with $k$ defects at positions $j_1,\dots,j_k$ by $\ket{j_1,\dots,j_k} = (\prod_{m=1}^k \hat{\sigma}^x_{j_m})\ket{vac}$. Using standard perturbation theory, the Ising ground state is expanded up to second order as
\begin{equation}
\begin{split}
    \ket{\Psi_0} = &\ket{vac} + \frac{1}{4h}\sum_i\ket{i,i+1}\\
    &+ \frac{1}{16 h^2}\sum_{j>i+1}\ket{i,i+1,j,j+1}\\
    &+ \frac{1}{8 h^2}\sum_{i}\ket{i,i+2} + \mathcal{O}(h^{-3}).
    \end{split}\label{eq:largeh}
\end{equation}
Notice that defects always appear in pairs, which is a crucial feature leading to entanglement enhancement.

Usually, an expansion like this is not particularly useful for many-body systems, as any finite truncation does not provide a good approximation of the true state. The way we will use Eq.~\eqref{eq:largeh} is to understand how measurements modify $\textit{all}$ perturbative orders collectively. As discussed below, this framework explains qualitatively all our numerical observations on the EE and fermionic negativity for $h<1$ and $h>1$.

Let us focus on $\mathcal{M}_\mathrm{up}$ first. When we apply $\ket{1_n}\bra{1_n}$ to $|\Psi_0\rangle$, some components of the expansion are filtered out, namely, all states with a difect at site $n$. We see from Eq.~\eqref{eq:largeh} that the leading $0$th-order state is left unchanged, while $\mathcal{O}(L^{n-1})$ states are removed from each order $n>0$. Since the population of each perturbative order is unaffected at leading order in $L$, the post-measurement state takes the form $|\Psi'\rangle=\ket{1_j}\ket{\varphi}$, where $\ket{\varphi}$ is a state of $(L-1)$ spins with analogous structure to $\ket{\Psi_0}$. 
Since the overlap $|\langle \Psi'|\Psi_0\rangle|\simeq 1$, we conclude that the projection represents a small perturbation to the hierarchical structure of the expansion.
Furthermore, the weights of the different states in the superposition are modified only slightly by the renormalization of the wavefunction. For this reason, quantum correlations can only decrease, as one spin has been factorized while no properties of the rest of the chain have changed significantly.

Crucially, a very different phenomenology is present for the protocol $\mathcal{M}_\mathrm{down}$, as applying $\ket{-1_n}\bra{-1_n}$ to $\ket{\Psi_0}$ alters the state remarkably. In this case, the projected state reads 
\begin{equation}
\begin{split}
    \ket{\Psi'} = &\frac{1}{4h}\left(\ket{n-1,n}+\ket{n,n+1}\right)\\
    &+ \frac{1}{16 h^2}\sum_{i\neq n,n-1,n-2}\ket{i,i+1,n-1,n}\\
    &+ \frac{1}{16 h^2}\sum_{i\neq n,n\pm 1}\ket{i,i+1,n,n+1}\\
    &+ \frac{1}{8 h^2}\left(\ket{n-2,n}+\ket{n,n+2}\right) + \mathcal{O}(h^{-3}).
    \end{split}\label{eq:largeh_projected}
\end{equation}
Therefore, the projection alters the structure of the state, shifting all orders by one. Most importantly, the populations of the various orders are renormalized. While the two leading orders of Eq.~\eqref{eq:largeh} contain one and $L$ states, respectively, those of Eq.~\eqref{eq:largeh_projected} contain twice as many (considering the leading power of $L$). As a consequence, more states participate to the superposition with relatively large amplitudes, and thus more entanglement is present. It is worth noting that the two new leading-order states of Eq.~\eqref{eq:largeh} form a Bell pair configuration for the spins on sites $n-1$ and $n+1$, further indicating an increase in entanglement. We stress, however, that entanglement enhancement cannot be reduced to the change of the leading-order, but rather it is an effect produced by the \emph{collective renormalization} of populations of the various orders. In fact, the overlap of the full $\ket{\Psi_0}$ with the leading-order state only is exponentially small in $L$, and thus it cannot impact the entanglement properties alone. 
When multiple down projections are performed, one may expect that the populations of lower orders increase exponentially in the number of measured sites, possibly leading to entanglement growth despite the decimation of the measured spins.

Finally, the Born rule protocol represents an intermediate case between forced up and down projections. Since $|\langle 1_j|\Psi_0\rangle|^2>|\langle -1_j|\Psi_0\rangle|^2$, entanglement enhancing jumps are rare, and the expected behavior is similar to the $\mathcal{M}_\mathrm{up}$ protocol.

\begin{figure*}[t]
    \centering
    \includegraphics[width=\textwidth]{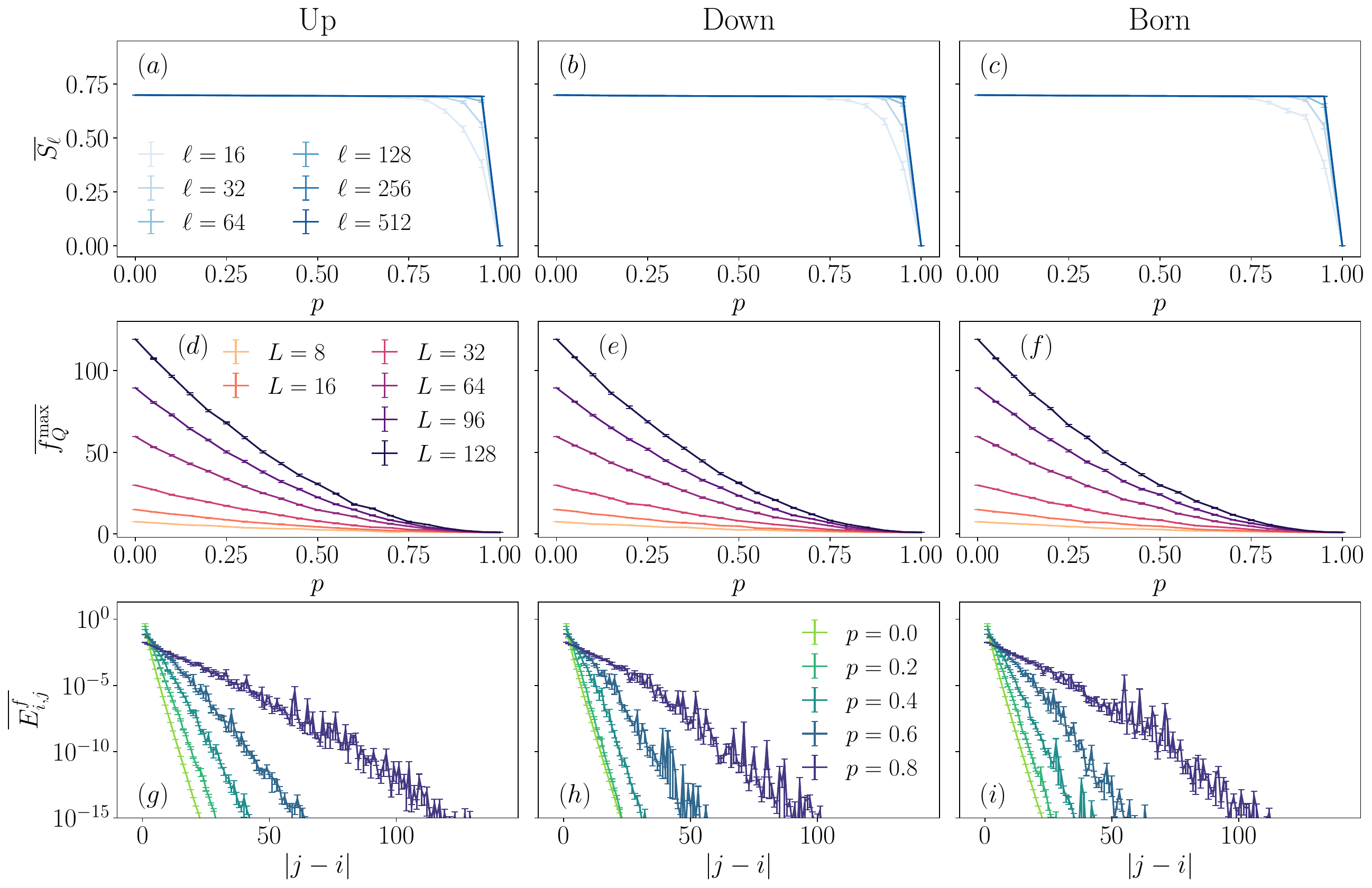}
    \caption{(a-c) EE, (d-f) maximal QFI density, and (g-i) pairwise FN of the perturbed GS of the quantum Ising chain with $h=0.5$. Left column: protocol $\mathcal{M}_\mathrm{up}$. Center column: protocol $\mathcal{M}_\mathrm{down}$. Right column: protocol $\mathcal{M}_\mathrm{Born}$. We adopt $L=1024$ for the EE and $L=512$ for the FN.}
    \label{fig:ordered}
\end{figure*}

\subsection{Ordered phase $h<1$}
Different entanglement behavior is observed in the ordered ferromagnetic phase, as we show in Fig.~\ref{fig:ordered} for the representative value of $h=0.5$~\footnote{We consider $L$ finite and then take the thermodynamic limit $L\to\infty$, following Ref.~\cite{wen2004quantum}. }. In the following, we consider the symmetry unbroken ground state, namely, with no net longitudinal magnetization $\langle \hat{\sigma}^x_j\rangle = 0$. In Fig.~\ref{fig:ordered} (a-c) we show the average EE. All projection protocol yield qualitatively similar results: thus, in contrast to the disordered phase, no enhancement is present for any $\mathcal{M}$. 
Additionally, for large subsystem sizes $\ell$, the EE appears to be approximately unaffected by the measurements, only decreasing when $p$ is close to $1$. 
As a consequence, the EE for any $p<1$ is qualitatively captured by the limit $p=0$, i.e., by $|\Psi_0\rangle$, provided the subsystem size is sufficiently large. 

We complement this analysis with the study of the QFI, reported in Fig.~\ref{fig:ordered} (d-f). Since the ordered phase features ferromagnetic long-range order, and we are considering a state with $\langle \hat{\sigma}^x_j\rangle = 0$, the correlator $C^{x,x}_{i,j}$ approaches a finite value for $|j-i|\to \infty$. As a consequence, the QFI density is extensive at $p=0$. We observe that this remains true for all $p>0$, and measurements renormalize the amplitude of the linear scaling. As for the EE, all protocols yield qualitatively similar results.

Last, Fig.~\ref{fig:ordered} (g-i) shows the two-spin FN.
Despite the absence of EE nor QFI enhancement, $\overline{E^f_{i,j}}$ decays exponentially with a length scale that grows with $p$, as in the case of $\mathcal{M}_\mathrm{down}$ in the disordered phase. 
Interestingly, this result applies to all protocols we consider, consistent with the results of the other witnesses.

Also in this case, we can develop an analytical understanding using perturbation theory. In the ferromagnetic phase, we expand the state around the symmetry-unbroken GHZ ground state for $h=0$, obtaining
\begin{equation}\label{ghz}
\begin{split}
    \ket{\Psi_0} = \,&\frac{\ket{+_1\dots+_L}+\ket{-_1\dots-_L}}{\sqrt{2}}\\ &+ \frac{h}{4}\sum_i \frac{\ket{+_1\dots-_i\dots+_L}+\ket{-_1\dots+_i\dots-_L}}{\sqrt{2}}\\&+\mathcal{O}(h^2),
\end{split}
\end{equation}
where $\hat{\sigma}^x\ket{\pm}=\pm\ket{\pm}$. 
States at order $n$ in the expansion present $n$ defects, corresponding to spin flips. Given that the spin states in Eq.~\eqref{ghz} are in the $x$ basis, both projectors $\ket{1_n}\bra{1_n}$ and $\ket{-1_n}\bra{-1_n}$ impact the state in similar ways. The difference only relies in the phases acquired by the projected wavefunction, because $\braket{1}{\pm} = 1/{\sqrt{2}}$ while $\braket{-1}{\pm} = \pm 1/\sqrt{2}$. As a consequence, it is not surprising to observe that the different protocols yield similar EE [cf. Fig.~\ref{fig:ordered} (a-c)], even though $\ket{\Psi_0}$ has a finite transverse magnetization and thus the jump probabilities are asymmetric.

To explain the behavior of entanglement with $p$, let us assume the case of $\mathcal{M}_\mathrm{up}$. For simplicity, let us consider the action of the projector $\ket{1_1}\bra{1_1}$ on the first lattice site. This yields a projected state $|\Psi'\rangle = \ket{1_1}\ket{\varphi}$, where
\begin{equation}
\begin{split}
    \ket{\varphi} = \,&\left(1+\frac{h}{2\sqrt{2}}\right)\frac{\ket{+_2\dots+_L}+\ket{-_2\dots-_L}}{\sqrt{2}}\\ &+ \frac{h}{4}\sum_{i=2}^L \frac{\ket{+_2\dots-_i\dots+_L}+\ket{-_2\dots+_i\dots-_L}}{\sqrt{2}}\\&+\mathcal{O}(h^2).
\end{split}
\end{equation}
Apart from subleading corrections to the amplitudes of each order, this state shares the same structure as $\ket{\Psi_0}$, with the difference that it involves only ${L-1}$ spins. 
As a consequence, the action of the projector approximately amounts to decimating a spin and mapping the remaining others to the Ising ground state with one less spin (and slightly modified parameters). Due to this chain renormalization, if we measure spin $n$, the entanglement between spins ${i<n}$ and ${j>n}$ after the projection is approximately the same as the entanglement between spins $i$ and $j-1$ before the projection.
This implies the growth of the average entanglement length scale shown in Fig.~\ref{fig:ordered} (d-f). Notice, however, that this mechanism simply shifts quantum correlations spatially, but does not grow their magnitude. As a consequence, as shown in Fig.~\ref{fig:ordered} (a-c), there can be no enhancement in the EE. For large $\ell$, $S_\ell$ is independent of $p$ because the quantum correlations that cross the bipartition are still intact, even though they may have been translated spatially by the measurements. For small $\ell$, instead, the entropy starts decreasing at large $p$, as most spins in the subsystem are decimated.

\begin{figure*}[t]
    \centering
    \includegraphics[width=\textwidth]{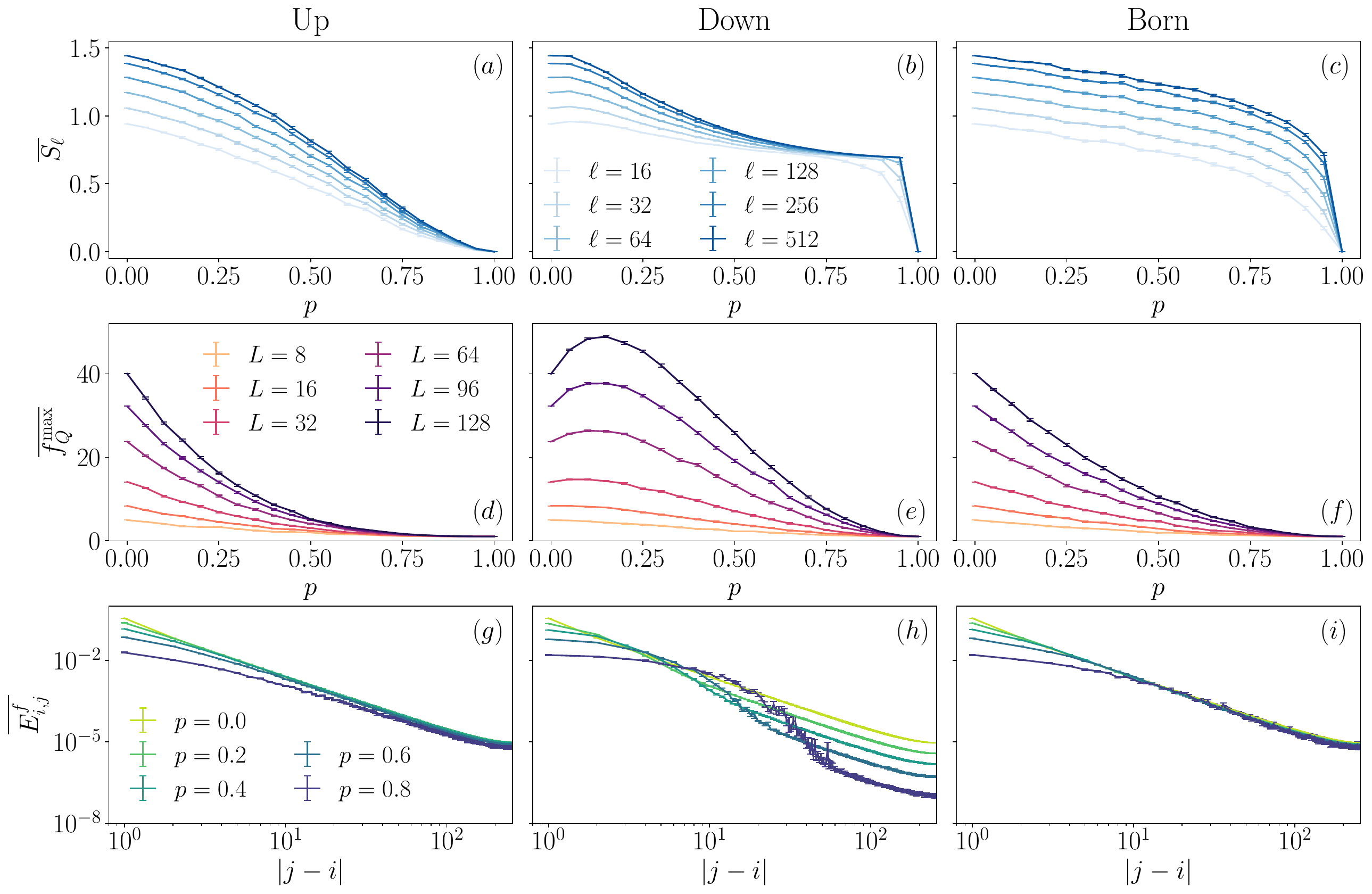}    
    \caption{(a-c) EE, (d-f) maximal QFI density, and (g-i) pairwise FN of the perturbed GS of the quantum Ising chain with $h=1$. Left column: protocol $\mathcal{M}_\mathrm{up}$. Center column: protocol $\mathcal{M}_\mathrm{down}$. Right column: protocol $\mathcal{M}_\mathrm{Born}$. We adopt $L=1024$ for the EE and $L=512$ for the FN.}
    \label{fig:crit}
\end{figure*}

\subsection{Critical point $h=1$}
Finally, in Fig.~\ref{fig:crit} we report the numerical results for the critical point $h=1$. 
As we demonstrate in Fig.~\ref{fig:crit} (a-c), overall bipartite entanglement decreases with $p$ for all protocols $\mathcal{M}$. 
In particular, $\mathcal{M}_\mathrm{up/Born}$ reproduce the results in the literature~\cite{lee2023quantum,weinstein2023nonlocality}, which are understood in terms of a defect perturbed field theory.
Interestingly, for $\mathcal{M}_\mathrm{down}$, the critical behavior is reminiscent to that of the off-critical phases, namely, the entropy appears to reach zero discontinuously for $p\to 1$.
For all protocols, we observe that at any given $p<1$ the EE maintains logarithmic dependence on $\ell$, with a central charge $c_\mathrm{eff}(p)$ renormalized by the external monitoring, consistently with previous results in the literature. Notice that for $\mathcal{M}_\mathrm{down}$ it may appear that the curves corresponding to different values of $\ell$ converge to a limiting line for $\ell\to\infty$ at large $p$ [see Fig.~\ref{fig:crit}(b)]. This fact would indicate a transition to an area law. However, performing a finite-size analysis, we investigated this possibility by estimating the central charge $c_\mathrm{eff}(p)$ using the largest values of $\ell$ available, and found that it reaches zero only at $p=1$.

Next, we analyze the QFI, summarizing our findings in Fig.~\ref{fig:crit} (d-f). 
While $\mathcal{M}_\mathrm{up}$ and $\mathcal{M}_\mathrm{Born}$ yield a qualitatively similar decrease with the measurement density $p$, for the $\mathcal{M}_\mathrm{down}$ protocol we observe enhanced multipartite entanglement.
In detail, a small density of projections $p$ can increase the QFI density. The peak position appears to shift towards larger $p$ as $L$ increases. Nevertheless, the effect of monitoring does not alter the scaling behavior of $f_Q$ on $L$, analogously to the EE.

Finally, Fig.~\ref{fig:crit}(g-i) highlights the pairwise negativity at the critical point. 
In this case, the decay is power law, and mirrors the divergence of the length scale of quantum correlations.
All protocols maintain the same behavior and do not affect the exponent of the power law. However, for $\mathcal{M}_\mathrm{down}$ this holds only asymptotically in the distance $|j-i|\to\infty$, whereas a distinct behavior arises at short distances within a region whose size grows with $p$.

Our numerical findings show that a finite density of measurements can give rise to both bipartite and multipartite entanglement enhancement, as well as to a renormalization of the length scale of quantum correlations.
In the next section, we describe this effect by means of a network toy model.

\section{Entanglement network toy model}
\label{sec:network}
Our results on the pairwise FN highlight that some measurement protocols can increase the typical length scale $\xi$ of quantum correlations. 
Local projections disentangle a spin from the chain, and consequently break some local correlations within the scale $\xi$. 
Nevertheless, if the projected spin is partially correlated with some other, new quantum correlations can be developed among these additional degrees of freedom. Motivated by our numerical results, we formulate a network toy model that captures the qualitative features of this phenomenon. It describes the action of measurements as cutting and sewing of quantum correlations, and, as we demonstrate below, it reproduces the key feature of entanglement enhancing, i.e., the renormalized average length scale of entanglement $\xi$. 
We anticipate that the model is not intended as a quantitative method, as it does not discriminate between different measurement protocols, nor it involves any notion of strength of entanglement.
Rather, its purpose is to provide a simplified description of those quantum projections that alter the state significantly (as in the cases of $h>1$ for $\mathcal{M}_\text{down}$ and $h<1$ for all protocols), and develop a qualitative understanding of the microscopic mechanism that leads to the average growth of $\xi$.

In our toy model, we describe quantum correlations between two spins at positions $i$ and $j$ as bonds in a ring network with $L$ vertices. The presence or absence of a bond is indicated by the binary variable $E_{i,j} = E_{j,i} =0,1$, whose values one and zero refer to the spins being entangled or not, respectively. This picture is a simplified version of the pairwise negativity (cf. Sec.~\ref{sec:maic}). Heuristically, this description corresponds to a 0th-order negativity spectrum, a pathological limit that measures only the rank of $\hat{\rho}^{\tilde{T_1}}$ (cf.~Sec.~\ref{sec:maic}).
We represent an initial state with finite entanglement length scale $\xi_0$ using the input state ansatz
\begin{equation}
    E_{i,j} =\begin{cases}
        1 \quad \text{if}\,\, |j-i|\leq\xi_0;\\
        0 \quad\text{otherwise.}
    \end{cases}
\end{equation}
Quantum projections act on the network as follows. Let us denote by $E_{i,j}$ and $E'_{i,j}$ the network configurations before and after the measurement. Suppose that the measured spin is at site $n$. First, since the projection disentangles it from the chain, we must have
\begin{equation}
    E'_{i,n} = 0 \quad \forall i.
\end{equation}
In addition, we assume that the degrees of freedom that were initially entangled with site $n$ develop new correlations by recoupling among themselves. (\emph{En passant}, we note this procedure resembles methods of strong disorder decimations, cf. Ref.~\cite{IGLOI_2005,Igl_i_2018}. )
In detail, for $i,j\neq n$ we impose 
\begin{equation}
    E'_{i,j} = \begin{cases}
        1 \quad\text{if}\,\, E_{i,n}=E_{j,n}=1;\\
        E_{i,j} \quad\text{otherwise.}
    \end{cases}
\end{equation}
In other words, the sub-network of spins initially coupled to $n$ becomes all-to-all connected, whereas the rest of the system remains unaffected. Notice that these rules will generally lead to multiple bonds for each site: this distinguishes our approach, for instance, from the network model presented in Ref.~\cite{nahum2020} where only pairwise links are allowed.

We formulated this recoupling scheme by investigating how the true negativity $E^f_{i,j}$ is affected by individual quantum measurements, and found that this mechanism works reasonably well for a binary representation of quantum correlations. Moreover, we point out that the previous rules guarantee the commutativity of measurements, which must hold since projectors acting on different sites commute.

\begin{figure}[t]
    \centering
    \includegraphics[width=0.85\columnwidth]{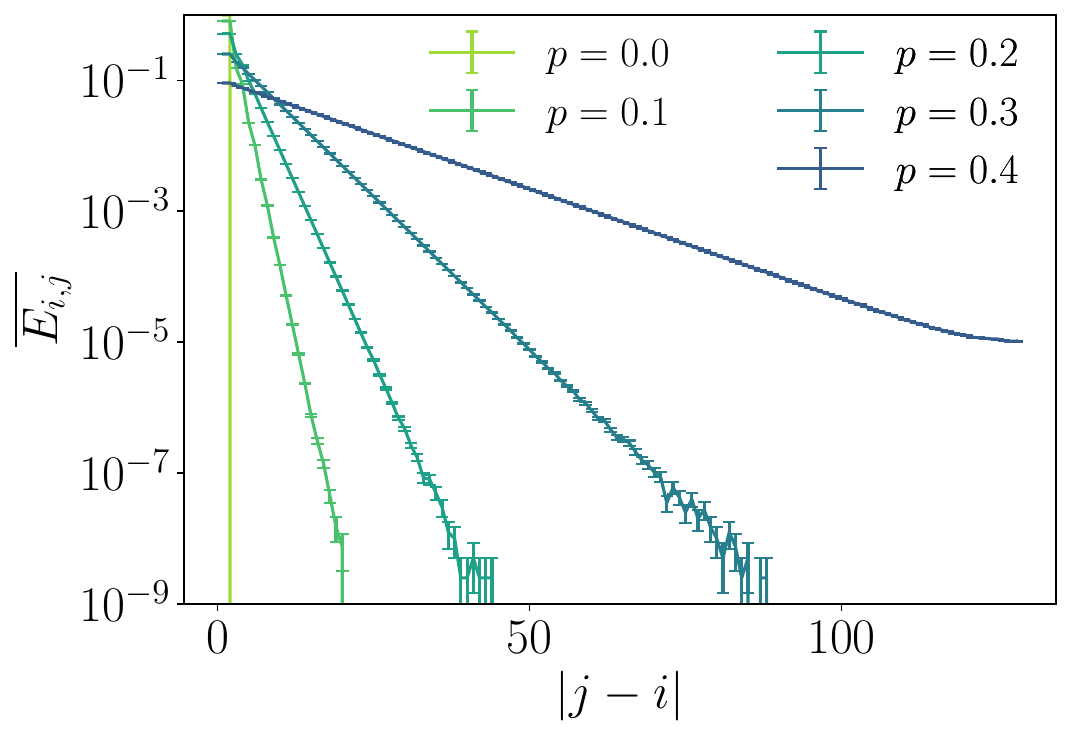}
    \caption{Average entanglement $\overline{E_{i,j}}$ of the toy model for $L=256$ and $\xi_0=2$. The entanglement estimator decays exponentially, with a lengthscale that increases with $p$. 
    }
    \label{fig:toy_model}
\end{figure}

We implement numerically the toy model as follows. Starting from an initial configuration with small $\xi_0$, we select randomly a fraction $p$ of the spins to project according to the previous network recoupling rule. We then repeat the procedure multiple times to compute the average $\overline{E_{i,j}}$. Figure~\ref{fig:toy_model} shows the results for $\xi_0=2$, representative of an area law correlated state. We observe that bonds remain short-ranged on average, as $\overline{E_{i,j}}$ decays exponentially, but the length scale increases with $p$. This reproduces the entanglement enhancement observed in Sec.~\ref{sec:numerics} for the pairwise FN.

\begin{figure}[t]
    \centering
    \includegraphics[width=\columnwidth]{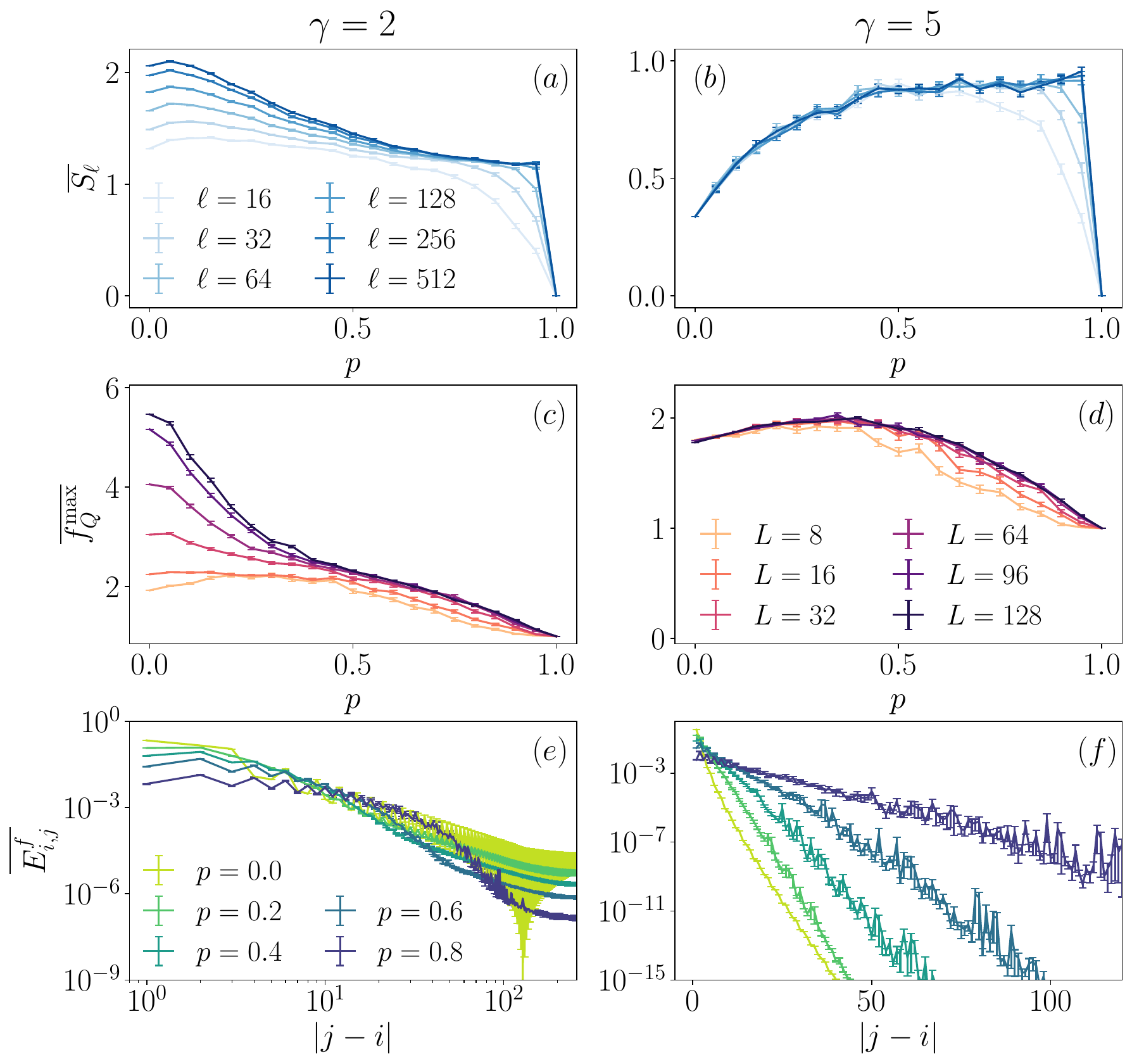}    
    \caption{(a-b) EE, (c-d) maximal QFI density, and (e-f) pairwise FN of the perturbed stationary state of the non-Hermitian Ising chain with $(h,\gamma)=(0.5,2)$ (left column) and $(h,\gamma)=(0.5,5)$ (right column), representative of other choices in the two phases}. The protocol used is $\mathcal{M}_\mathrm{up}$. We adopt $L=1024$ for the EE and $L=512$ for the FN.
    \label{fig:nh}
\end{figure}

\section{Stability test for the no-click limit of weak-measurement dynamics}
\label{sec:mipt}
Some works in the literature on monitored quantum dynamics discuss measurement-induced phase transitions in a particular limit, the so-called no-click limit, which corresponds to non-Hermitian Hamiltonian models that can be approached theoretically to achieve analytical results~\cite{biella2021manybodyquantumzeno,gopalakrishnan2021,paviglianiti2023multipartiteentanglementin,turkeshi2023entanglementandcorrelation,zerba2023measurement,granet2023,barch2023}. While it provides a simplified scenario to investigate the phenomenon, it is now understood that the no-click limit is unable to fully capture the more complex physics found in stochastic monitored dynamics, as the two manifest differences in phase boundaries~\cite{turkeshi_erratum,paviglianiti2023multipartiteentanglementin,legal2024} and critical exponents~\cite{leung2023}. In the following, we leverage the framework of projected ensembles to investigate whether these discrepancies can be at least partially reproduced by perturbing the no-click limit with a layer of measurements, which is the simplest way to incorporate a key ingredient of the stochastic dynamics. Concretely, we repeat the analysis of the measurement-altered Ising chain for a different initial state $\ket{\Psi_0}$, which this time is the no-click stationary state. Since the monitored evolution does not include Born rule measurements, we focus only on a single forced measurement protocol corresponding to the quantum jumps of the stochastic dynamics.

Before discussing the phenomenology of the measurement-altered non-Hermitian model, let us provide a brief overview on the no-click limit. The evolution of a system undergoing unitary dynamics and measurement is described by a stochastic Schr\"odinger equation, which embeds the intrinsic randomness arising from the external monitoring as a non-unitary time-dependent Hamiltonian term. In the case of a weak monitoring protocol, which for example may represent quantum optical setups where the system is coupled to a photodetector, a commonly studied model is the quantum jump equation
\begin{equation}\label{sse}
    d\ket{\Psi_t} = -i\hat{H}_\mathrm{NH}dt\ket{\Psi_t}+\sum_{j}dN_{j,t}\left(\frac{\ket{1_j}\bra{1_j}}{|\langle 1_j |\Psi_t\rangle|}-1\right)\ket{\Psi_t},
\end{equation}
which involves a non-Hermitian Hamiltonian
\begin{equation}\label{ising_nh}
    \hat{H}_\mathrm{NH} = \hat{H} -i\frac{\gamma}{4}\sum_j\left(\hat{\sigma}_j^z-\langle\hat{\sigma}_j^z\rangle_t\right),
\end{equation}
plus sudden local projections implemented by $\ket{1_j}\bra{1_j}$. In the previous expression, $\hat{H}$ is the Hermitian Hamiltonian of the system without measurements. The random variables $dN_{j,t}=0,1$ are increments of independent Poisson processes with averages $\overline{dN_{j,t}}=\gamma dt |\langle 1_j |\Psi_t\rangle|^2$, and set the statistics of random quantum jumps. A brief derivation of Eq.~\eqref{sse} can be found in App.~\ref{a:sse_derivation}, as well as in Ref.~\cite{turkeshi2022entanglementtransitionsfrom}. Within this framework, sudden quantum jumps occur only to the state $\ket{1_j}$ with positive magnetization, whereas the complementary measurement outcome yields an infinitesimal yet time-continuous projection to $\ket{-1_j}$, implemented by the non-Hermitian Hamiltonian. 

The properties of the non-Hermitian Hamiltonian have been investigated in Refs.~\cite{turkeshi2023entanglementandcorrelation,biella2021manybodyquantumzeno}. $\hat{H}_\mathrm{NH}$ drives the system toward a stationary state, fixed by the many-body eigenstate of $\hat{H}_\mathrm{NH}$ with the largest imaginary eigenvalue. The dynamics produced by the Hamiltonian alone is referred to as no-click limit, as no sudden jump takes place.
The non-Hermitian model features two distinct phases. For $|h|<1$ and $\gamma<\gamma_c(h) = 4\sqrt{1-h^2}$, the EE of the stationary state scales logarithmically with $\ell$, whereas outside this region it saturates to a constant. A relationship between the measurement-induced transition, where no post-selection is present, and the non-Hermitian model has been recently proposed in Ref.~\cite{turkeshi2022entanglementtransitionsfrom}.
However, it is not clear in which setups and up to what extent the non-Hermitian Hamiltonian describes quantitatively the properties of the system.

Let us now investigate how quantum jumps perturb the results of the no-click limit. In particular, we are interested in establishing whether the no-click entanglement scaling properties are robust against measurements. For instance, in principle it might occur that a logarithmic-law entangled state is unstable against measurements, which degrade it to an area law. Choosing $\ket{\Psi_0}$ to be the stationary state of $\hat{H}_\mathrm{NH}$, we perturb it with a layer of forced up projections corresponding to the protocol $\mathcal{M}_\mathrm{up}$ of Eq.~\eqref{prob_density_up}, as this is the jump direction introduced by the stochastic Schr\"odinger equation. As mentioned above, such a stationary state is just the eigenstate of $\hat{H}_\mathrm{NH}$ with largest imaginary part, and is thus  thus accessible analytically by solving the model exactly. For $|\Psi_0\rangle$ in the area law phase, all results on the entanglement witnesses are completely analogous to the case of the disordered Ising ground state (cf. Fig.~\ref{fig:disordered}). This is highlighted in the right column of Fig.~\eqref{fig:nh}, which shows numerical results for $\gamma > \gamma_c(h)$. The only difference between the Hermitian and non-Hermitian cases is that the roles of $\mathcal{M}_\mathrm{up}$ and $\mathcal{M}_\mathrm{down}$ are swapped. This is easily understood in terms of the transverse magnetization of $|\Psi_0\rangle$. For the Hermitian Ising chain at $h>1$, the ground state has $\langle \hat{\sigma}_j^z\rangle>0$, and quantum jumps produce entanglement enhancement only for measurements that oppose this magnetization, i.e., the protocol $\mathcal{M}_\mathrm{down}$. In contrast, the stationary state of Eq.~\eqref{ising_nh} for $\gamma>0$ has net magnetization $\langle \hat{\sigma}_j^z\rangle<0$, and thus the enhancing protocol is $\mathcal{M}_\mathrm{up}$. In particular, forced up projections applied to the stationary state result in enhanced EE and QFI, as well as in an enhanced length scale $\xi$ of the pairwise FN.
Regarding the logarithmic phase of the non-Hermitian model, we present our results in the left column of Fig.~\eqref{fig:nh}, which considers $\gamma<\gamma_c(h)$. The behaviors of the various witnesses are qualitatively similar to those of the critical Ising ground state of Fig.~\ref{fig:crit}, provided swapping $\mathcal{M}_\mathrm{up}$ and $\mathcal{M}_\mathrm{down}$ as discussed above. 
The only notable difference is that the QFI density of Fig.~\ref{fig:nh}(c) does not feature a peak, as opposed to Fig.~\ref{fig:crit}(h). This is not surprising, as the Hermitian and non-Hermitian states $\ket{\Psi_0}$ considered do not feature an identical entanglement structure. For instance, in both cases the QFI density at $p=0$ scales as a power law, but the exponents are not the same.

Overall, the behaviors of the non-Hermitian model for $\gamma<\gamma_c(h)$ and $\gamma>\gamma_c(h)$ closely resemble what we found in the Hermitian case for $h=1$ and $h>1$, respectively. The scaling laws of all entanglement witnesses are unaffected by the measurement-altering scheme, proving that a single layer of measurements is not enough to affect them drastically and induce a crossover to the physics of the full monitored dynamics. We then conclude that the phenomenology generated by Eq.~\eqref{sse} requires the interplay between random measurements and unitary evolution, and cannot be recovered by a perturbed no-click limit.

\section{Conclusions}
\label{sec:conclusion}
This manuscript discusses the entanglement properties of the measurement-altered Ising chain for different measurement protocols, measurement densities, and initial states. 
As the main result, our numerical analysis finds that suitable measurement outcomes produce entanglement enhancement, thus demonstrating that local measurements may increase the system entanglement. The principle beyond this effect is based on the monogamy of entanglement, and is exemplified in the perturbative analysis of Sec.~\ref{sec:numerics} and in the toy model discussed in Sec.~\ref{sec:network}. This challenges the conventional interpretation of measurement-induced phase transitions as a competition between entanglement creation, due to scrambling dynamics, and destruction, due to measurements. 
In our investigation, enhancement is always intensive, in the sense that measurement-altering does not change the (non-)critical nature of the state.

The measurement-altered framework provides insights into the relationship between the properties of non-Hermitian Hamiltonians and measurement-induced phases.
Recalling that projections in forced directions arise naturally in quantum jump stochastic Schr\"odinger equations, we studied the stability of the non-Hermitian stationary state. The critical or non-critical nature of the starting state persists at all finite measurement densities, demonstrating its stability under this protocol. Any discrepancy between non-Hermitian physics and measurement-induced transitions must be looked for in the dynamical nature of latter.

It would be interesting to investigate the occurrence of entanglement enhancement in different models with distinct entanglement structure, as well as with interactions~\cite{xing2023interactions}. For instance, a natural question may be what happens to measurement-altered systems in two or more dimensions, where area law EE corresponds to a scaling $S_\ell$. Another possible direction of investigation regards choosing different projection operators, e.g., local projectors in the longitudinal direction of spontaneous symmetry breaking, or corresponding to generalized measurements. Lastly, one may expect to observe a non-trivial phenomenology in presence of entangling measurements, such as projections onto neighboring Bell pairs. These issues are left for future studies.

\section{Acknowledgements}
\label{sec:acknowledgements}

A.S. would like to acknowledge support from PNRR MUR project PE0000023-NQSTI and Quantera project SuperLink. X.T. and M. S. acknowledge support from the ANR grant “NonEQuMat” (ANR-19-CE47-0001). 

\appendix

\section{Numerical implementation}\label{a:implementation}
In this Appendix we discuss all details about the numerical implementation of our study. First, we briefly introduce the Jordan-Wigner transformation used to map spins to fermions. We then explain how to implement measurements within the formalism of Gaussian states. Finally, we describe how to implement the measurement protocol in practice.

The quantum Ising chain of Eq.~\eqref{ising} can be mapped to a BCS fermionic Hamiltonian using the Jordan-Wigner map. We introduce the fermionic operators $\hat{c}_j$ and $\hat{c}\daga_j$ by defining
\begin{subequations}
\begin{equation}
    \hat{\sigma}^+_j = \frac{\hat{\sigma}^x_j+i\hat{\sigma}^y_j}{2} = e^{i\pi\sum_{i=1}^{j-1} \hat{n}_i} \hat{c}_j,
\end{equation}
\begin{equation}
    \hat{\sigma}^-_j = \frac{\hat{\sigma}^x_j-i\hat{\sigma}^y_j}{2} = e^{i\pi\sum_{i=1}^{j-1} \hat{n}_i} \hat{c}\daga_j,
\end{equation}
\end{subequations}
where $\hat{n}_i = \hat{c}\daga_i\hat{c}_i$. Equation~\eqref{ising} is mapped (up to an additive constant) to 
\begin{equation}\label{hamiltonian_fermions}
\begin{split}
    \hat{H} = &- \sum_{j=1}^{L-1} \left(\hat{c}\daga_j\hat{c}_{j+1} + \hat{c}\daga_j\hat{c}\daga_{j+1} + \mathrm{h.c.}\right) \\
    &+ (-1)^{\hat{N}} \left(\hat{c}\daga_L\hat{c}_1
    + \hat{c}\daga_L\hat{c}\daga_1 + \mathrm{h.c.}\right) + 2 h\sum_{j=1}^L \hat{n}_j,        
\end{split}
\end{equation}
where $\hat{N}=\sum_j\hat{n}_j$. Due to the $\mathbb{Z}_2$ symmetry of the model, the parity $(-1)^{\hat{N}}$ is a conserved quantity and the Hamiltonian can be diagonalized separately in each symmetry sector. In our investigation, we consider the ground state of the Ising chain, which lies in the even parity sector for even $L$. We thus assume $(-1)^{\hat{N}}=1$, and we obtain a fermionic chain with anti-periodic boundary conditions. This can be diagonalized by moving to momentum space through the definition
\begin{equation}\label{momentum_transformation}
    \hat{d}_k = \frac{e^{-i\pi/4}}{\sqrt{L}}\sum_j e^{-i k j} \hat{c}_j,
\end{equation}
and performing a standard Bogoliubov transformation. This leads to the diagonal form
\begin{equation}
    \hat{H} = \sum_k \epsilon_k \hat{\gamma}\daga_k \hat{\gamma}_k,
\end{equation}
where the sum runs over $k = \pm \frac{2n-1}{L}\pi$, $n=1,\dots, L//2$, and the quasiparticle excitation energies are given by $\epsilon_k = 2\sqrt{1-2h\cos k + h^2}$. The ground state of the model is the quasiparticle vacuum state $\ket{\Psi_0} = \otimes_{k>0}\ket{\text{vac}_k}$, where $\hat{\gamma}_k\ket{\text{vac}_k} = \hat{\gamma}_{-k}\ket{\text{vac}_k} = 0$. Denoting by $\ket{0}_k$ the vacuum of $\hat{d}_{\pm k}$, we have explicitly
\begin{equation}\label{vacuum}
    \ket{\text{vac}_k} = \frac{(\epsilon_k-2\cos k +2 h)\ket{0_k} + 2\sin k\,\hat{d}\daga_k\hat{d}\daga_{-k}\ket{0_k}}{\sqrt{2 \epsilon_k (\epsilon_k-2\cos k +2 h)}}.
\end{equation}

Being an eigenstate of a quadratic Hamiltonian, the ground state $\ket{\Psi_0}$ is Gaussian. This implies that it is fully described by its correlation matrices $C_{m,n} = \langle \hat{c}_m\hat{c}\daga_n\rangle$ and $F_{m,n} = \langle \hat{c}_m\hat{c}_n\rangle$. The correlation matrices for $\ket{\Psi_0}$ are easily obtained from Eqs.~\eqref{vacuum} and~\eqref{momentum_transformation}. It can be shown that the projective measurements we consider in Sec.~\ref{sec:maic} preserve the Gaussianity of the state. As a consequence, we can study our problem by just keeping track of $C$ and $F$ and updating them appropriately when we measure the state. The update equations are derived as follows. Let $\ket{C}$ and $F$ be the correlation matrices of a Gaussian state $\ket{\Psi}$, and let $C^{(1,j)}$ and $F^{(1,j)}$ be those of the projected state $\ket{\Psi^{(1,j)}} = \hat{c}_j\hat{c}_j\daga\ket{\Psi}/\sqrt{C_{j,j}}$ (the denominator normalizes the state); this corresponds to a measurement of $\hat{\sigma}_j^z$ with outcome $+1$. The new correlation matrices are given by
\begin{subequations}
\begin{equation}
    C^{(1,j)}_{m,n} = \frac{\langle \hat{c}_j\hat{c}_j\daga \hat{c}_m\hat{c}\daga_n \hat{c}_j\hat{c}_j\daga\rangle}{C_{j,j}},
\end{equation}
\begin{equation}
    F^{(1,j)}_{m,n} = \frac{\langle \hat{c}_j\hat{c}_j\daga \hat{c}_m\hat{c}_n \hat{c}_j\hat{c}_j\daga\rangle}{C_{j,j}},
\end{equation}
\end{subequations}
and they can be expressed in terms of $C$ and $F$ using Wick's theorem to break down the expectation values involving six fermionic operators. The result of this calculation reads
\begin{subequations}
\begin{equation}
    C^{(1,j)}_{m,n} = C_{m,n} - \frac{C_{m,j}C_{j,n}}{C_{j,j}} + \frac{F_{m,j}F\daga_{j,n}}{C_{j,j}} + \delta_{j,m}\delta_{j,n},
\end{equation}
\begin{equation}
    F^{(1,j)}_{m,n} = F_{m,n} - \frac{C_{m,j}F_{j,n}}{C_{j,j}} + \frac{F_{j,m}C_{n,j}}{C_{j,j}}.
\end{equation}
\end{subequations}
Similarly, for a measurement with negative outcome the projected state is $\ket{\Psi^{(-1,j)}} = \hat{c}\daga_j\hat{c}_j\ket{\Psi}/\sqrt{1-C_{j,j}}$, and the update rules are found to be
\begin{subequations}
\begin{equation}
\begin{split}
    C^{(-1,j)}_{m,n} = &C_{m,n} + \frac{(\delta_{m,j}-C_{m,j})(\delta_{j,n}-C_{j,n})}{1-C_{j,j}}\\
    &- \frac{F_{m,j}F\daga_{j,n}}{1-C_{j,j}} - \delta_{j,m}\delta_{j,n},
\end{split}
\end{equation}
\begin{equation}
    F^{(-1,j)}_{m,n} = F_{m,n} - \frac{(\delta_{m,j}-C_{m,j})F_{j,n}}{1-C_{j,j}} + \frac{F_{j,m}(\delta_{n,j}-C_{n,j})}{1-C_{j,j}}.
\end{equation}
\end{subequations}

The state projection protocol is implemented as follows. For each lattice site $j$ we pick a random number $r\in[0,1]$. If $r>p$ we leave it unaffected, otherwise we perform a measurement. For $\mathcal{M}_\text{up}$ and $\mathcal{M}_\text{down}$ the choice of the projector to apply is predetermined, whereas for $\mathcal{M}_\text{Born}$ we compute $\langle \hat{\sigma}_j^z\rangle)$ on the current state and we choose by applying the Born rule. After running over all sites, we measure the entanglement witnesses of the resulting state. The full procedure is iterated multiple times to collect statistics and evaluate their averages.

\section{Derivation of the quantum jump equation}\label{a:sse_derivation}

In this Appendix, we derive the quantum jump equation of Eq.~\eqref{sse}, introduced in Sec.~\ref{sec:mipt}. We consider the quantum Ising chain of Eq.~\eqref{ising}, and we assume a weak measurement protocol for the spins. A generalized measurement with $M$ possible outcomes can be defined by assigning $M$ Kraus operators $\hat{A}_m$ that satisfy the identity $\sum_m\hat{A}^\dag_m\hat{A}_m = \hat{\mathds{1}}$~\cite{wiseman1996,ahnert2005,brun2002}. When acting on an initial state $\ket{\Psi_t}$, the post-weak-measurement is
\begin{equation}
    \ket{\Psi_{t+dt}^{(m)}} = \frac{\hat{A}_m\ket{\Psi_t}}{\sqrt{\bra{\Psi_t}\hat{A}^\dag_m\hat{A}_m\ket{\Psi_t}}},
\end{equation}
with the $m$-th outcome picked with probability  $p_m = \bra{\Psi_t}\hat{A}^\dag_m\hat{A}_m\ket{\Psi_t}$. Considering each spin in the chain independently, we evolve the state from $t$ to $t+dt$ according to the measurement defined by
\begin{equation}
    \begin{split}
        \hat{A_j}^{(-1)} &=\ket{-1_j}\bra{-1_j} +\sqrt{1-\gamma dt} \ket{1_j}\bra{1_j} ;\\
        \hat{A_j}^{(1)} &= \sqrt{\gamma dt} \ket{1_j}\bra{1_j},
    \end{split}
\end{equation}
where the parameter $\gamma$ controls the rate of quantum jumps. The operator $\hat{A}_j^{(-1)}$ implements an infinitesimal deviation of the starting state towards $\ket{-1_j}$, and occurs with large probability. In contrast, $\hat{A}_j^{(1)}$ abruptly projects the state locally onto $\ket{1_j}$ with a small probability $\propto \gamma dt$.

For each site $j$ and at each time $t$, let us introduce a binary random variable $dN_{j,t}=0,1$ with $\overline{dN_{j,t}} = \gamma dt |\braket{1_j}{\Psi_t}|^2$. By construction, the stochastic process $dN_{j,t}=0,1$ has the same statistics of quantum jumps, and can thus be used to model them. In detail, when $dN_{j,t}=1$, we apply the projector $\hat{A}_j^{(1)}$ to the state. In this case, the post-measurement state is given by
\begin{equation}
    \ket{\Psi_{t+dt}^{(1,j)}} = \frac{\ket{1_j}\braket{1_j}{\Psi_t}}{|\langle 1_j |\Psi_t\rangle|}.
\end{equation}
In contrast, when $dN_{j,t}=0$ for all $j$, the state evolves infinitesimally due to the Hamiltonian $\hat{H}$, as well as the action of the Kraus operator $\hat{A}_j^{(-1)}$, yielding
\begin{equation}
    \ket{\Psi_{t+dt}^{(-1)}} \approx \left[1-\frac{\gamma dt}{2}\sum_j\left(\ket{1_j}\bra{1_j}-|\langle 1_j |\Psi_t\rangle|\right)\right]\ket{\Psi_t}
\end{equation}
at leading order in $dt$. Processes involving two simultaneous jumps on different sites are subleading in $dt$, and are thus neglected. The full evolution of the state is given by the combination of the two
\begin{equation}\label{qje}
\begin{split}
    \ket{\Psi_{t+dt}} = &e^{-i\hat{H}dt}\bigg[\prod_j(1-dN_{j,t}) \ket{\Psi_{t+dt}^{(-1)}}\\
    &+ \sum_j dN_{j,t} \ket{\Psi_{t+dt}^{(1,j)}}\bigg],
\end{split}
\end{equation}
where we have included the unitary Hamiltonian evolution. Finally, the quantum jump equation of Eq.~\eqref{sse} is obtained from Eq.\eqref{qje} by keeping only the leading order term in $dt$ for both the deterministic and the stochastic parts, cf.~\cite{paviglianiti2023multipartiteentanglementin}. The trajectory where no jump occurs ($dN_{j,t}=0$ for all $j,t$) is given by the non-Hermitian evolution, Eq.~\eqref{sse}. 

\bibliography{biblio}

\end{document}